\documentclass{ws-mpla}
\usepackage[super]{cite}
\usepackage{graphicx}

\newcommand{\ben}{\begin{enumerate}}\newcommand{\een}{\end{enumerate}}

\newcommand{\om}{\omega}

\renewcommand{\kappa}{\varkappa}
\newcommand{\q}{\mathbf{q}}
\renewcommand{\a}{\mathbf{a}}\renewcommand{\c}{\mathbf{c}}
\renewcommand{\k}{\mathbf{k}}

\newcommand{\n}{\mathbf{n}}
\newcommand{\Nn}{\mathbf{N}}

\newcommand{\tfi}{\tilde{\phi}}
\newcommand{\Tr}{{\rm Tr}}
\newcommand{\M}{{\cal M}}
\begin{document}

\markboth{Irina Pirozhenko}
{On Finite Temperature Casimir Effect for Dirac Lattices}

%%%%%%%%%%%%%%%%%%%%% Publisher's Area please ignore %%%%%%%%%%%%%%
\catchline{}{}{}{}{}
%%%%%%%%%%%%%%%%%%%%%%%%%%%%%%%%%%%%%%%%%%%%%%%%%%%%%%%%%%%%%%%%%%%

\title{ON FINITE TEMPERATURE CASIMIR EFFECT \\ FOR DIRAC LATTICES
}

\author{\footnotesize IRINA PIROZHENKO}

\address{Bogoliubov Laboratory of Theoretical Physics, Joint Institute for Nuclear Research \\ and \\
Dubna State University,
Dubna, 141980,
Russia\\
pirozhen@theor.jinr.ru}

\maketitle

\pub{Received (Day Month Year)}{Revised (Day Month Year)}

\begin{abstract}
We consider polarizable sheets modeled by a lattice of delta function potentials.  The Casimir interaction of two such lattices is calculated at nonzero temperature.  The heat kernel expansion for periodic singular background is discussed in relation with the high temperature asymptote of the free energy.
\end{abstract}

\ccode{PACS Nos.: 11.10.Wx, 12.20.Ds}

\section{The System}	
In recent years, we have witnessed the discoveries of ever new unusual properties of two-dimensional materials, such as 2d electron gas, graphene, or other monoatomic layers. In the present note we deal with a system which mimics two parallel monoatomic layers and discuss van der Waals and Casimir forces between them.  Two-dimensional rectangular lattices of delta function potentials separated by a distance $b$ are considered. Our objective is to evaluate the vacuum energy of a scalar field in  the background of these lattices and compute the finite temperature corrections.

A periodic delta potentials are a well studied in quantum mechanics, the simplest case being the Kronig-Penney model
 (`Dirac comb') \cite{Albeverio1988}. In more than one dimension a Hamilton operator with a delta function potential is not self adjoint,  however  a self-adjoint extension may be defined. For a Laplace operator with tree-dimensional $\delta$ function, $\Delta_a=\Delta+a \delta^{3}(x)$, the self-adjoint extension was analyzed\cite{solo99-541-461}  in the QFT context. Close relation of self-adjoint extension and zero range potential approach with regularization and renormalization was traced~\cite{jack91,bord15-91-065027}.

The scattering on a single two-dimensional lattice 3D delta functions was considered
 in Ref.~\citen{bord15-91-065027} for scalar and electromagnetic field.
 In Ref.~\citen{bord17-95-056017} the scattering approach was used to obtain  the separation dependent part of the vacuum energy for a scalar field in the presence of two parallel lattices. In the notations of this paper, the lattice sites of two rectangular 2D lattices, (A) and (B), are given by 3D vectors
\begin{equation} \vec{a}^{\rm \,A}_{\bf n}=\left(\begin{array}{c}{\bf a}_{\bf n}+{\bf c} \\ b\end{array}\right),
    \quad   \vec{a}^{\rm \,B}_{\bf n}=\left(\begin{array}{c}{\bf a}_{\bf n}  \\0\end{array}\right), \quad
    {\bf a}_{\bf n}=a\left(\begin{array}{c} n_1 \\ n_2\end{array}\right),
\end{equation}
${\bf a}_{\bf n}$ are $2D$ vectors in $(x,y)$-plane, $n_1$ and $n_2$ are integers, $a$ is the lattice spacing,
$b$ is the separation,  ${\bf c}$ is the displacement vector in $(x,y)$-plane.

The wave equation for a scalar field $\phi(\vec{x}),$
\begin{equation} \left(-\omega^2-\Delta+g\sum_{\bf n}\left(
    \delta^{(3)}(\vec{x}-\vec{a}^{\rm \,A}_{\bf n})+\delta^{(3)}(\vec{x}-\vec{a}^{\rm \,B}_{\bf n})\right)
              \right)\phi(\vec{x})=0,
\label{2.2}
\end{equation}
endowed with two lattices of three dimensional delta functions is not well defined, and  $g$,  having the dimension of length, should be regarded as a bare coupling\cite{bord15-91-065027}.

In the scattering approach the separation dependent part of the vacuum energy is given by so called `TGTG'-formula, see for example \cite{Klich_2009},
\begin{equation}
E_0=\frac{1}{2\pi}\int_0^\infty d\xi \, \Tr\ln(1-\M(i\xi,\vec{x},\vec{x}')),  \quad  \M= T_A G_0  T_B G_0,
\label{2.1}
\end{equation}
where  $\xi$ is the imaginary frequency, $\om=i\xi$, and the trace is taken with respect to $\vec{x}$.   The free Green's function is denoted by
\begin{eqnarray}
 G_0(\vec{x}-\vec{x}')
 %   &=&\int\frac{d^3k}{(2\pi)^3}\,\frac{e^{i\vec{k}(\vec{x}-\vec{x}')}}{\om^2-\k^2-k_3^2+i0}.
%\nn\\
 &=& \int\frac{d^2\k}{(2\pi)^2}\,\frac{e^{i\vec{k}(\vec{x}-\vec{x}')+i\Gamma |x_3-x_3'|}}{2i\Gamma(\k)}, \quad \Gamma(\k)=\sqrt{\om^2-k^2+i0},
\label{2.6}
\end{eqnarray}
and the $T$ operators describe scattering on the lattices $A$ and $B$
\begin{eqnarray}
 T_{ A,B}(\vec{x},\vec{x}')= \sum_{\n,\n'}\delta(\vec{x}-\vec{a}^{\rm \,A,B}_{\n})
    \Phi^{-1}_{\n,\n'}     \delta( \vec{a}^{\rm \,A,B}_{\n'}-\vec{x}') .
\label{2.5}
\end{eqnarray}
Here $\Phi^{-1}_{\n,\n'}$ is the inverse matrix to
$ \Phi_{\n,\n'}=\frac{1}{g}\delta_{\n,\n'}-G_0(\vec{a}_\n-\vec{a}_{\n'})$ with
 diagonal elements defined after the renormalization\cite{bord15-91-065027} of the coupling $g$ so that  $\Phi_{\n,\n}=1/g$.

Due to translational invariance with respect to the lattice step  the momentum $\k$  may  be
split into quasi momentum and integer part, $\k=\q+\frac{2\pi}{a}\Nn$. The infinite momentum integration is replaced by  $\int d^2\k=\int d^2\q\sum_\Nn$ with the components of $\q =( q_{1} ,q_2)$ restricted to
$-\pi/a \le {q}_{1,2}<\pi/a$. One can derive $\Tr\ln(1-\M)$  in (\ref{2.1}) expanding the logarithm in powers of $\M(i\xi)$, which after some transformations\cite{bord17-95-056017} appear to be  diagonal with respect to $q$.  Consequently  the expression (\ref{2.1}) can be written in a Lifshitz-like form. For the vacuum energy per lattice cell we get
\begin{equation}
E_0 =\frac{1}{2}\int_0^\infty \frac{d\xi}{\pi}\, a^2\int\frac{d^2\q}{(2\pi)^2} \ln\left(1-|h(i\xi,\q)|^2\right),
\label{2.1.17}
\end{equation}
with
\begin{equation} \quad h(\om,\q) = \frac{1}{a^2}\sum_\Nn\,
    \frac{e^{i \Gamma(\k)b+i\frac{2\pi}{a}\Nn\c}}{2i\Gamma(\k)\tfi(\k)}, \quad \tilde{\phi}(\k) = \frac{1}{g}-\frac{1}{4\pi}{\sum_\n}' \frac{1}{|\a_\n|}e^{i \om |\a_\n|+i \k \a_\n}
\label{2.1.14}
\end{equation}
%$$\tilde{\phi}(\k) = \frac{1}{g}-\frac{1}{4\pi}J_1(\om,\k), \quad J_1(\om,\k)={\sum_\n}' \frac{1}{|\a_\n|}e^{i \om |\a_\n|+i \k \a_\n} $$
%
(in the primed sum the term with $\n=0$ is dropped).
For week coupling $g$  it corresponds to  the vacuum energy of parallel plates with ``reflection coefficient" $r(\omega,\k)=g/(2 a^2)(\omega^2-\k^2)^{-1/2}$.

The general formula (\ref{2.1.17}) obtained in Ref.~\citen{bord17-95-056017} for the vacuum energy of two lattices at zero temperature will be used to find finite temperature corrections.

\section{Finite Temperature}
%%%%%%%%%%%%%%%%%%%%%%%%%%%%%%%%%%%%%%%%%%%%%%%%%%%%%%
To find the finite temperature corrections the Matsubara formalism is used. The free energy is given by the sum over   Matsubara frequencies $i\xi\to \xi_n=2\pi T n$,
 \begin{equation} \mathcal F=\frac{T}{2}\sum_{n=-\infty}^{\infty} {\rm Tr} \ln
\left(1-{\cal M}(\xi_n)\right)=\frac{T a^2}{2}\sum_{n=-\infty}^{\infty} \int\frac{d^2\q}{(2\pi)^2} \ln\left(1-|h(\xi_n,\q)|^2\right) .
\label{3.1}
\end{equation}
%Here we define $\Gamma_T=\sqrt{(2\pi n T)^2+k^2}$.

We define three temperature regions  with respect to the parameters of the model, namely the lattice spacing $a$ and the separation of two lattices $b$. Low temperature corresponds to $Ta, Tb \ll 1$.  Medium temperature meets the inequality $T a < 1< Tb$. This area requires numerical studies. High temperature obeys the condition $1\ll Ta \ll T b$.
Here it is worth mentioning that when  $b\ll a$ the problem reduces to the interaction of two $\delta$ sources.

%\subsection{Low temperature}
 For low  temperature, $Ta, Tb \ll 1$, the Abel-Plana formula may be used
  \begin{equation} \mathcal F=E_0+\mathcal F_T, \quad
   \mathcal{F}_T=
\frac{i}{2}\int_{-\infty}^\infty\frac{d\xi}{2\pi}\ n_T(\xi)
 \, \mbox{Tr}  \left[\ln(1-{\cal M}(i\xi))-\ln(1-{\cal M}(-i\xi))\right].
\label{3.2}
\end{equation}
 Here $n_T(\xi)=(\exp(|\xi|/T)-1)^{-1}$ is the Boltzman factor.
At low temperatures
the integral in (\ref{3.2}) is determined by small $\xi$.
After the  development around $\xi=0$, each power of $\xi$ adds a power of $T$,
\begin{eqnarray}
 &&  \mathcal{F}_T=
\frac{1}{\pi}\int\limits_{-\infty}^\infty\frac{d\xi}{\exp(\frac{|\xi|}{T})-1}[{\cal M}_1 \xi + {\cal M}_3 \xi^3 +\dots]={\cal M}_1 \frac{\pi T^2}{6} + {\cal M}_3 \frac{\pi^3 T^4}{15} +\dots,
\label{3.3}\\
&& {\cal M}_1=-\Tr\frac{{\cal M}'}{1-{\cal M}}, \quad {\cal M}_3=-\Tr\left\{\frac{{{\cal M}'}^3}{3(1-{\cal M})^3}+\frac{{\cal M}' {\cal M}''}{2(1-{\cal M})^2}+\frac{ {\cal M}'''}{6(1-{\cal M})}\right\} \dots \nonumber.
\end{eqnarray}

%\subsection{High temperature}
At high temperature the leading contribution is given by the zeroth therm of the Matsubara sum (\ref{3.1}) which is proportional to the derivative of the spectral zeta function, $\zeta'(0)$, of considered system.  The subleading terms are expressed through  the coefficients  of the heat kernel  expansion $
K(t)=\sum_{j}e^{-\lambda_j^2t}\sim(4\pi t)^{-3/2}\sum_{n=0}^{\infty}t^{n/2} a_{n/2}$,  $t\to0$.
Thus the entire high temperature expansion aquires the form,
\begin{eqnarray}
\mathcal{F}(T) &\simeq&-\frac{T}{2}\zeta'(0)+a_0\frac{T^4}{\hbar^3}\,
\frac{\pi^2}{90}-\frac{a_{1/2}}{4\pi^{3/2}}\,\frac{T^3}{\hbar^2 }
\zeta_{\text {R}}(3) -\frac{a_1}{24}\frac{T^2}{\hbar} +\dots.
%+\frac{a_{3/2}}{(4\pi)^{3/2}}\,T\,\ln\frac{\hbar}{T}.
\label{3.4}
\end{eqnarray}

The heat kernel in the spectral problem~(\ref{2.2}) with two parallel delta lattices $A$ and $B$ is given by an integral equation,
%\begin{equation}
%K(x,y;t)=K_0(x,y;t)+\int_{0}^t ds \int_{V}d^3 z \, K_0(x,z;t-s)V(z)K(z,y;s)
%\label{3.5}
%\end{equation}
\begin{eqnarray}
K(x,y;t)&=&K_0(x,y;t)
+g\int_{0}^t ds \sum_{i=A,B} \sum_{\mathbf{n}}  K_0(x,a^{\rm \,i}_{\mathbf{n}};t-s)K(a^{\rm \,i}_{\mathbf{n}},y;s) ,
\label{3.5}
\end{eqnarray}
 with the
free heat kernel $ K_0(x,y;t)=(4\pi t)^{-3/2} e^{-\frac{(x-y)^2}{4t}}$.  The integral equation (\ref{3.5})  can be iterated\cite{gave86-19-1833,Bordag1999}. After taking the trace  one arrives at $K(t)=K_0(t)+K_{(1)}(t)+K_{(2)}(t)+K_{(3)}(t)...$, where
\begin{eqnarray}
K_0(t)=\frac{1}{(4\pi t)^{3/2}}V, \quad
K_{(1)}(t)=\frac{2 N^2 g}{(4\pi)^{3/2}t^{1/2}}, \dots.
\label{3.6}
\end{eqnarray}
Here $N$ is the number of the  lattice points.   Unfortunately, starting from $g^2$ term we come across divergent integrals which may be regularized by point-splitting. Similar problem was discussed in Ref.~\citen{Bordag1999} . It is worth  comparing~(\ref{3.6}) with the exact solution for the heat kernel trace of the operator with a single three-dimensional $\delta$-function,  obtained in Ref.~\citen{solo99-541-461}
\begin{equation}
K(t)=\frac{1}{(4\pi t)^{3/2}}+\frac{1}{2}e^{\frac{16 \pi^2}{g^2}t}\left[1-\Phi\left(\frac{4\pi}{g}\sqrt{t}\right)\right], \quad \Phi(z)=\frac{2}{\sqrt{\pi}}\int_0^z dx e^{-x^2}.
\label{3.7}
\end{equation}
The small $g$ expansion of (\ref{3.7})  coincides  in its  leading terms with $K_0(t)$ and $K_{(1)}(t)$,  up to  volume  and arear factors.
Thus, from  (\ref{3.6})   the leading heat kernel coefficients  $a_0$ and $a_{1/2}$  may be extracted.

\section{Conclusion}
We considered the Casimir effect for two-dimensional lattices of delta functions at zero and finite temperature.  Our  approach is based on the $TGTG$ formula, with  $T$-operators derived in terms of lattice sums.  The generalization to finite temperature is tricky but straightforward. Here scaling properties of the system and relations between various limiting cased proved to be useful at low and high temperatures. The heat kernel of the Laplace operator with double delta lattice potential was analyzed in relation with the high temperature asymptote of the free energy.  Medium temperature region is left for numerical study.

\section*{Acknowledgments}
The author is thankful to the Organizers of 4th Symposium on the Casimir effect. She also acknowledges fruitful discussions with Michael Bordag.

\end{document}